# Information-guided optimization of image-based sensorless adaptive optics methods

**BIWEI ZHANG, MARTIN J. BOOTH, AND QI HU**

*Department of Engineering Science, University of Oxford, Oxford OX1 3PJ, UK*

**Abstract:** Adaptive optics (AO) are reconfigurable devices that compensate for wavefront distortions or aberrations in optical systems such as microscopes, telescopes and ophthalmoscopes. Aberrations have detrimental effects that can reduce imaging quality and compromise scientific information. Sensorless AO methods were introduced to correct aberrations without a separate wavefront sensor, inferring wavefront-related information directly from phase-diverse sample images. Most sensorless AO control systems, although effective and flexible to use, were operated based on empirical experience with suboptimal performance. In this paper, we introduced a Fisher information-based analysis framework to provide information-guided method optimization. Results suggested that our framework can effectively improve the accuracy and efficiency of different sensorless AO methods. The framework is not specific to any AO method or imaging modality and has the potential to benefit a wide range of applications.

**Introduction**

Microscopes suffer from wavefront distortions introduced by inherent limitations of optics, mismatched refractive indices and inhomogeneous sample structures. The detrimental effects of such optical aberrations reduce imaging quality and compromise scientific information. Adaptive optics (AO) uses reconfigurable optical elements to actively compensate for aberrations by introducing suitable wavefront modulations [1-4]. As there are many challenges to implementing wavefront sensor-based AO in microscopes, sensorless AO methods have often been used due to their flexibility and versatility [5-20]. In sensorless AO, aberrations are inferred from acquired phase-diverse images, rather than measured directly. The basis of most sensorless AO methods is to introduce known system disturbances while acquiring measurements of the image plane (or focal volume). By assuming the sample structure stays unchanged during the AO process, any change to the image properties results from the introduced disturbances that embed phase information. The required wavefront correction is then deduced from the measurement variations and the known introduced disturbances [21,22].

A common choice of the introduced system disturbance is phase modulation by AO [5-22]. Although different sensorless AO methods were developed following slight variations of the working principles, for many of those, method settings, such as the phase modulation magnitude and shape, were designed heuristically through empirical experience without systematic and objective optimization. Such approaches result in suboptimal method design, for which input data are not optimally conditioned. This results in prolonged iterations – requiring more image acquisitions – that are not desirable for many practical applications.

There is a direct relationship between the introduced phase modulation sequence and the phase information encoded in the images. However, the relationship is not obviously explicit. Practical factors such as the illumination power, sample brightness, noise, background level, and out-of-focus structures may all affect the detectable signals and, in turn, affect the phase information extraction. It is thus very difficult to determine objectively how phase modulations should be introduced to better condition the detected data.

In this paper, we present a Fisher information-based analysis to assess objectively how these interdependent factors jointly contribute to the extraction of phase information from microscope measurements. This provides the basis for a systematic approach to optimization of the phase modulation strategy that could better condition the image data for phase extraction.

As shown in Fig. 1, We computed the Cramér–Rao Lower bound from the Fisher information by assuming an unbiased estimator and hence made our analysis independent of any specific aberration estimation algorithm for a fair and more general study. We also extended the analysis by adopting features of specific AO algorithms to make the analysis more practically relevant. Simulated results showed that method settings optimized through the Fisher information analysis led to better correction accuracy under the same simulated imaging conditions. This work demonstrated a versatile framework that can provide valuable insights into AO method design for a wide range of imaging scenarios.

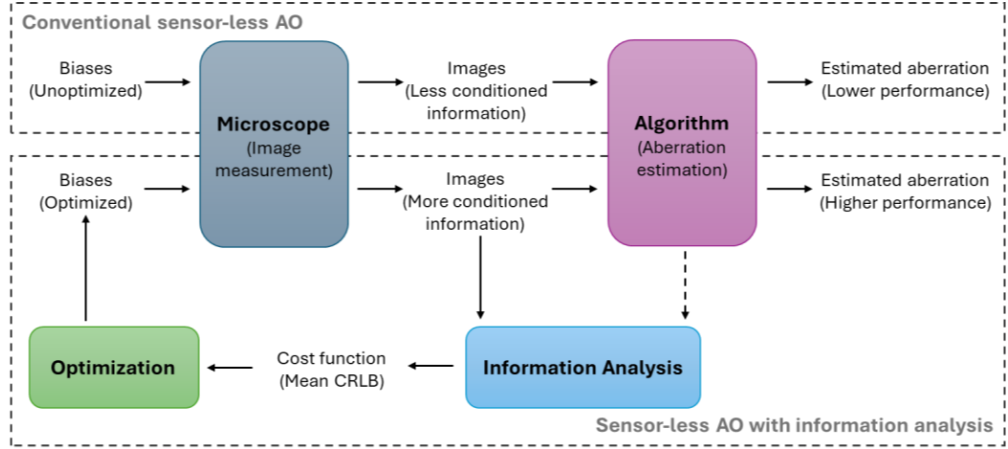

Fig. 1. A pipeline diagram shows how our information-guided analysis is used to optimize a sensorless AO method. We built a digital twin for a microscope and the synthesized images are used for information analysis to optimize the input biases for better-conditioned data acquisition.

**Method**

*Fisher information and Cramér–Rao lower bound*

Fisher information is a statistical measure of the amount of information carried by an observable variable $X$ about an unknown yet contributing parameter $\Theta$ [23]. It is thus suitable for the current application, as sensorless AO can be interpreted as the derivation of unknown phases from image measurements. To measure the variance (hence precision) of an estimator to $\Theta$ from observations $X$, the Cramér–Rao Lower bound (CRLB) was computed from the Fisher information. The CRLB sets the lower bound of the variance, and hence the highest possible accuracy, of the estimator [24].

Fisher information $I_X(\Theta)$ is defined as

$$I_X(\Theta) = E\left[\left(\frac{\partial}{\partial \Theta}\log f(X;\Theta)\right)^2\right], \qquad (1)$$

where $f(X;\Theta)$ is the probability density function of the observation $X$ conditioned on the value of $\Theta$. $E[\cdot]$ is the expected value.

Loosely speaking, Fisher information $I_X(\Theta)$ can be interpreted as measuring the changes of the distribution of $X$ with respect to an incremental change of $\Theta$. A larger change to the $X$ distribution resulting from a change of $\Theta$ means that $X$ carries more information about $\Theta$.

The Cramér–Rao Lower bound was defined as an inverse to Fisher information.

$$CRLB_{\hat{\theta}} = I_X(\Theta)^{-1}. \qquad (2)$$

In this work, $X$ is not always a single measurement but may be a collection of photo-detector readings, such as an image or a stack of images. For the purpose of this demonstration, we assumed that each observation was measured independently and followed a Poisson distribution. $\Theta$ is a vector space that consists of multiple parameters. The parameter space included both parameters to be estimated, such as aberration coefficients, as well as interdependent parameters, such as the object structure, the illumination power, wavelengths, the scanning range and pixel size, and objective numerical aperture.

A more general equation to encompass this multi-dimensional analysis can be written as

$$[I_X(\Theta)]_{l,m} = \sum_q \left[I_{X_q}(\Theta)\right]_{l,m} = \sum_q E\left[\left(\frac{\partial}{\partial \theta_l}\log f(X_q;\Theta)\right)\left(\frac{\partial}{\partial \theta_m}\log f(X_q;\Theta)\right)\right], \quad (3)$$

Where $\theta_l, \theta_m \in \Theta$ and $X_q \in X$. The CRLB of each estimator $\hat{\theta}_k$ was further calculated from the Fisher information matrix by:

$$CRLB_{\hat{\theta}_k} = [I_X(\Theta)^{-1}]_{k,k}. \quad (4)$$

Note that both $\theta_l$ and $\theta_m$ are parameters of the complete vector space $\Theta$. $\hat{\theta}_k$ form a subset $\hat{\Theta}_k \in \Theta$ consisting of those parameters that are to be estimated.

### Adaptive microscope numerical simulation

For illustration of the novel method, a specific implementation in a multiphoton fluorescence adaptive microscope was chosen for modelling. A numerical model of the multiphoton microscope was designed for the purpose of the analysis demonstration; this was based upon a scalar Fourier optics model [25], which is suitable for efficient modelling the effects of aberrations in microscopes. In most sensorless AO methods for adaptive microscopy, the phase information was deduced from a sequence of acquired images, each with a different predetermined phase modulation (often referred to as "bias aberrations"). The choice of these bias aberrations is a key aspect of the sensorless AO method design as they influenced how the phase information was encoded and conditioned. In this paper, we focused our discussion on deriving optimized bias aberrations and considered how the information content analysis can lead to more accurate and efficient sensorless AO method design.

For demonstration purposes, we limited the system aberration to contain only the first $K$ Zernike modes (different settings for each analysis will be defined separately in the results section). The total phase of the wavefront in the illumination pupil, $\phi$, for each collected biased image can be written as

$$\phi = \sum_k^K (\theta_k + a_k)Z_k, \quad (5)$$

where $\theta_k$ is the decomposed system aberration coefficient of Zernike mode $Z_k$ to be estimated, $a_k$ is the introduced bias aberration coefficient of Zernike mode $Z_k$. The resultant multiphoton point spread function (PSF) can be written as

$$PSF = \left[\left(FT(Pe^{i\phi})\right)\left(FT(Pe^{i\phi})\right)^*\right]^\beta, \quad (6)$$

where $FT(\cdot)$ is the two-dimensional Fourier transform; $P$ is a unit circular pupil function, where the region outside of this pupil has amplitude zero, and within the pupil is one; $i$ is the imaginary unit; $\beta$ is the order of nonlinearity, and $*$ is the complex conjugate. In this numerical model, the pixel size was set such that two pixel-widths matched the full-width-half-maximum of the PSF.

The multiphoton fluorescence microscope is an incoherent imaging system, so the observed image $X$ of an object with fluorescence distribution $O$ can be written as

$$X = \text{Poisson}\{\alpha(PSF \otimes O) + b\}, \quad (7)$$

where $\otimes$ is a two-dimensional convolution; $\alpha$ is a value indicating the effective brightness of the signal collection that is related to a number of factors including the sample brightness, illumination power, dwell time, fluorescence emission, and photo-detection dynamic range; $b$ is a value indicating the potential issues of background and signal from out-of-focus planes; $X$ is then computed following a Poisson distribution, Poisson$\{\cdot\}$. To model the imaging of 3-D extended objects, for computational simplicity we used an approximate model in which the focal volume was sampled at five equally spaced axial planes, and the spacing of any two adjacent planes was equal to the diffraction-limited axial full-width-half-maximum of the effective multiphoton PSF (e.g., $0.8 \mu m$ in our simulation of the two-photon microscope), with the central plane placed at the PSF centre. Each of the five 2-D planes of the 3-D PSF was convolved (x-y plane) with the corresponding 2-D plane of the 3-D object structure (where the object structure was also sampled at the same five equally spaced axial planes as the focal volume) and the resultant five 2-D images were summed in the axial dimension to compute one 2-D image.

When applying the analysis in different demonstrations as shown in the results section, $X$ may be computed as variations to the above definition depending on the input measurements to different sensorless AO methods. For the analysis in results section 1, the measurement $X$ was computed when $O$ was a point-like object and pixels in each image were summed to generate the total intensity input. For the analysis in results section 2 and 3, the measurements $X$ were images of object $O$ when $O$ was selected to be a point-like object or volumetric extended objects, depending on the different analyses. For sensorless AO methods with $N$ number of different biases and hence $N$ number of images $X$, each image was generated with the same system aberration, $\theta_k$, but different bias aberration, $a_k$. All the pixels in the resultant $N$ number of images $X$ were treated as independent and summed when calculating the Fisher information, as defined in Equation (3).

*Optimization process*

The purpose of the analysis in this paper is to derive the best set of bias aberrations $a_k$ that leads to the statistically most accurate estimation of $\theta_k$. To derive optimized estimators to $\Theta_k$, joint optimization through minimization of $CRLB_{\hat{\theta}_k}$ was conducted. A loss function was defined as:

$$Loss = \sqrt{\sum_k CRLB_{\hat{\theta}_k}}. \tag{8}$$

The optimization process used a framework similar to that used for artificial neural network training. To implement the optimization, all the $N$ bias aberrations were first randomly initialized. Simulated examples were randomly generated with different system aberrations (within a defined range) to form a dataset. The size of the dataset varied for different practices but was not smaller than 8,000; it was selected to have a sufficient coverage of the analysis range to ensure a stable convergence. Different ranges of system aberrations, in terms of root mean square (RMS) values, were chosen for different analyses and specified in the results section. All the random aberrations were generated following a uniform distribution in an n-sphere as defined in [26]. This means that any random combination of these five modes within the selected aberration RMS range had an equal opportunity to be generated. An Adam gradient descent optimizer was applied to minimize the mean loss over the dataset by optimizing the $N$ bias aberrations jointly [27]. To avoid memory leaks, mini-batch optimization was employed so that each optimization step was not based on the whole dataset but on a randomly selected subset [28]. The hyperparameters of the optimization (such as the learning rate, optimizer, initializer, number of steps, batch size, etc.) were finely adjusted (for detailed settings of the hyperparameters, please see the supplementary information). The implementation was programmed with Tensorflow in Python and ran on a GPU (Nvidia GeForce RTX 3070).

**Results**

*1. Optimized bias amplitude for intensity-based parabolic fitting methods*

The first application of the Fisher information analysis was to derive the optimal bias amplitude for a parabolic maximization method. Specifically, we used total image intensity as the sensorless AO correction feedback (observation $X$) and $2N+1$ parabolic fitting as the estimation algorithm. This is one of the most common sensorless AO methods used in a multiphoton microscope, where for each of the $N$ selected modes, a positive and a negative amount of the bias amplitude were introduced by an AO into the microscope system with a corresponding measurement taken. In addition, an extra measurement was taken when no phase modulation was introduced. A total of $2N+1$ measurements were used to correct $N$ modes. For each mode, two corresponding biased measurements together with the common non-biased measurement were each computed by summing all pixel values. The three metric readings were fitted to a parabola to extrapolate the metric peak and its corresponding optimal coefficient. For full details of this method, please refer to [5]. Two imaging modalities, a two-photon (2-P, Fig. 2a and 2b) and a three-photon (3-P, Fig. 2c and 2d) microscope, were included for analysis. The initial aberration was generated consisting of random combinations of five Zernike modes (astigmatism: $Z_5$, $Z_6$; coma: $Z_7$, $Z_8$; and spherical: $Z_{11}$). The RMS value of the random aberration was not larger than 0.5 rad (Fig. 2a and 2c) and 1 rad (Fig. 2b and 2d). This aberration range was chosen to ensure that the input aberrations were within the range of validity of the parabolic approximation to the intensity metric. For each mode, a range of different bias amplitudes was used to compute the mean CRLB loss and the mean estimation error of the coefficient for comparison.

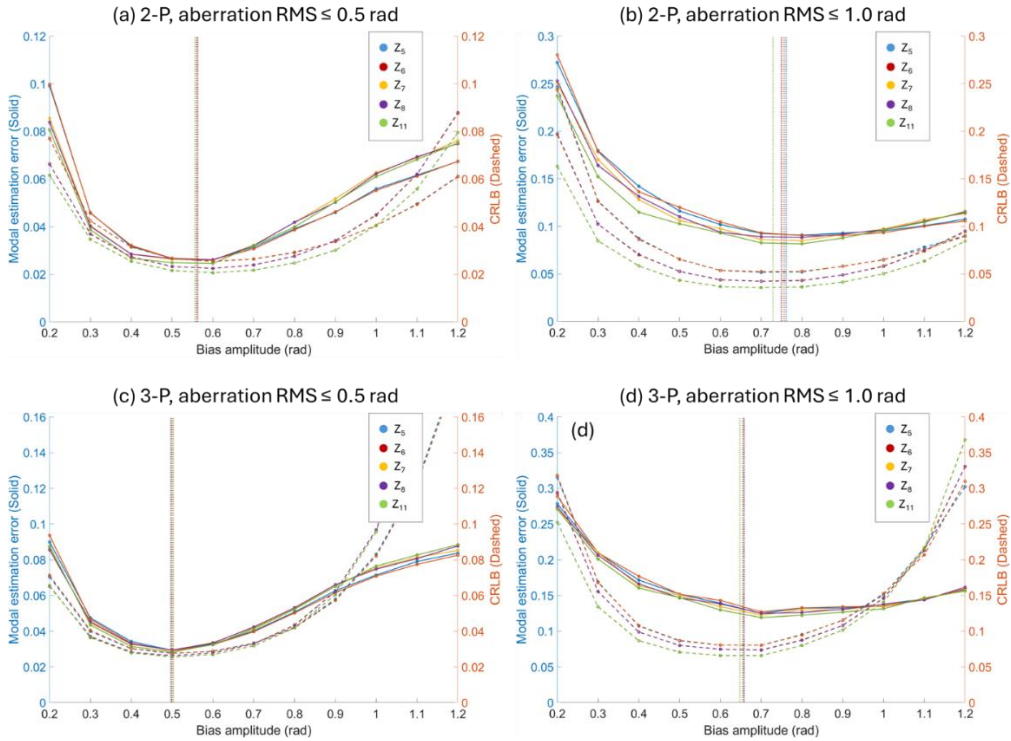

Fig. 2. Plots of the mean CRLB loss (dotted curve) comparing to simulated mean residual error for each mode estimated by $2N+1$ parabolic fitting algorithm (solid curve) when varying the bias amplitude. The initial aberrations were randomly generated with RMS phase not larger than (a, c) 0.5 rad and (b, d) 1.0 rad. The top row (a, b) simulated a 2-P microscope and the bottom row (c, d) simulated a 3-P microscope. In each plot, the bias amplitudes corresponding to the lowest mean CRLB loss were plotted in vertical dashed lines.

Figure 2 shows the simulated results averaged from 2000 tests. In Fig. 2a, 2b and 2c, the CRLB curves follow the general trends of the simulated residual errors after $2N+1$ parabolic fitting correction process. The optimized bias amplitudes derived from minimizing the CRLB loss (vertical dotted lines) also corresponded approximately to the minimum of the residual error curve. One interesting observation is that the two astigmatism modes (blue: $Z_5$ and red $Z_6$) always follow each other closely for both the CRLB curve and simulated residual error curve; similar for the two coma modes (yellow: $Z_7$ and purple: $Z_8$). This is expected as the two pairs of modes have similar shapes and only differ in orientations. The spherical mode: $Z_{11}$ for 2-P analysis generally has the smallest optimized bias amplitude; this was also reported in the previous work [5]. Another observation is that comparing Fig. 2a and 2b, the optimized bias amplitude has increased when we have aberrations with larger RMS, and this is also expected and reported in the previous work [29].

In Fig. 2d, when the maximal RMS of the random aberration has increased to 1 rad for a 3-P microscope, the CRLB loss curves start to deviate from the computed residual errors. The optimized bias amplitudes from the CRLB are offset from the minima of the residual error curves. However, as the residual error curve is very flat (and no significant variation within the bias range of 0.6 to 1.2 rad), the minima of the curves are not well defined and the CRLB derived optimum is also within this range. This is expected as the 3-P microscope has a higher non-linearity and hence a narrower response curve compared to a 2-P microscope. The validity of the parabolic fitting algorithm when correcting aberrations in a 3-P microscope is therefore only guaranteed over a smaller aberration range.

*2. Optimized bias modes for better-conditioned phase information*

In this demonstration, we derived optimal bias modal shapes that can better condition phase information in the images. The motivation behind this analysis is the emergence of machine learning-based sensorless AO methods in the literature [30-34]. The great computational power brought by neural networks allowed better use of image information, correcting wavefront as few as two images. However, most supervised learning-based methods still rely on heuristic design procedures for the AO image acquisition process. In [34], some analyses suggested that astigmatism was the best bias mode choice from a set of low-order aberration modes. In this section, we performed a more fundamental analysis of this mode choice, based upon information of whether a thin planar or thick volumetric sample was imaged during the AO correction process in a 2-P microscope. For the purpose of demonstration, the initial aberration was generated consisting of random combinations of seven Zernike modes (astigmatism: $Z_5$, $Z_6$; coma: $Z_7$, $Z_8$; trefoil: $Z_9$, $Z_{10}$; spherical: $Z_{11}$) with the RMS value smaller than 3 rad.

Figure 3 shows results from the analysis of Fisher information in two different imaging scenarios - imaging (i) point-like objects and (ii) extended volumetric objects. As shown in Fig. 3a, two derived optimal bias aberrations have shapes very similar to (i) defocus for point-like objects, and (ii) astigmatism for extended volumetric objects. For the volumetric objects, the defocus mode was removed from the analysis, given that it would correspond to focusing to a different plane of the 3-D structure. Figure 3b shows the statistical results over 2,000 tests of the mean CRLB loss and mean residual error. The residual error was computed in this case through a machine learning-based sensorless AO algorithm (for details, please see the supplementary information). The simulation shows that the optimized set of modes consistently provided the lowest residual errors compared to individual Zernike-mode biases.

One interesting observation is that when comparing Fig. 3b(i) and 3b(ii), coma ($Z_7$, $Z_8$) and trefoil ($Z_9$, $Z_{10}$) modes perform much differently for the point-like object and volumetric object simulation. The reason of the poor performance in the planar object simulations is because with an ideal, perfectly in-focus 2-D plane, PSF deformations due to opposite signed odd-row modes (e.g. astigmatism: $Z_5$, $Z_6$) are not distinguishable and their signs are ambiguous. On the other hand, when imaging volumetric samples, opposite signed odd-row modes have different PSF deformations at the-out-of-focus planes, which make them differentiable and non-ambiguous (for details of this discussion, please see the supplementary information).

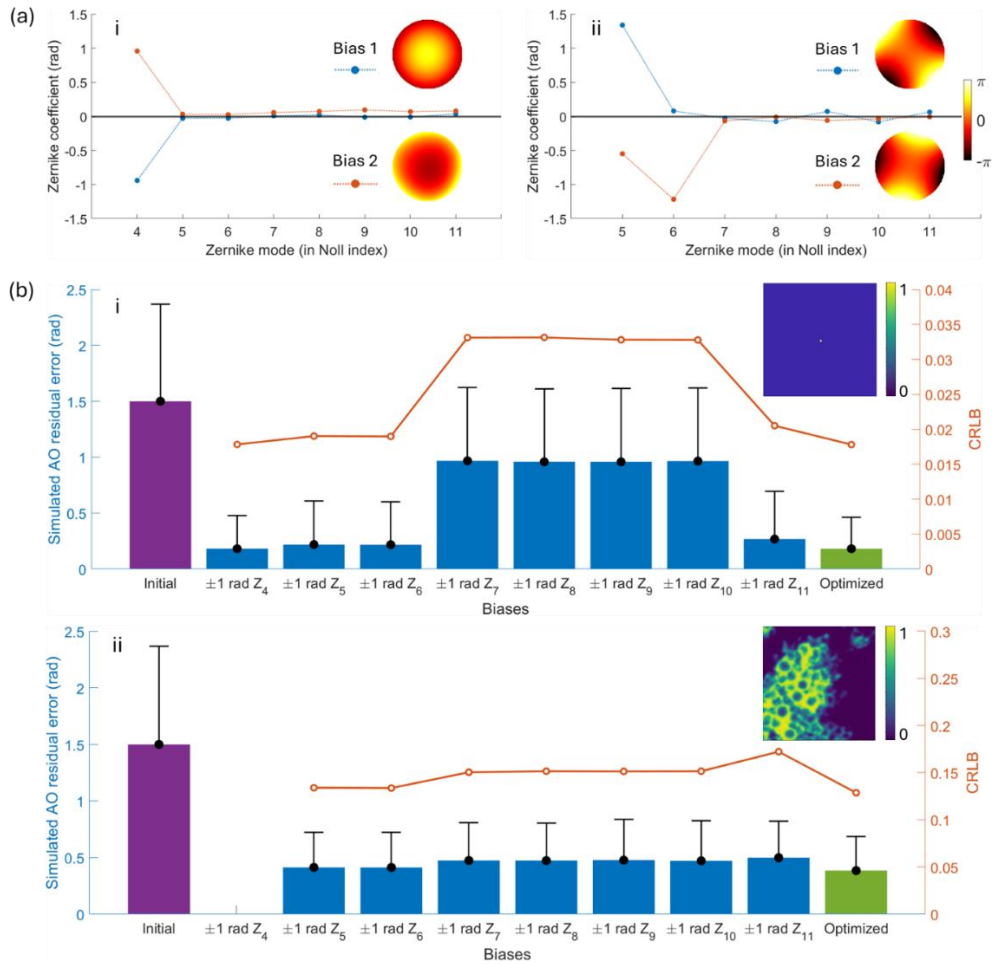

Fig. 3. (a) Fisher information analyses derived optimized biases when imaging (i) point-like objects and (ii) extended volumetric objects. The derived two optimal bias modes were decomposed, and their modal components were plotted as the red and blue data points. The insets showed the resultant shapes of the optimized bias modes. (b) Simulated sensorless AO mean residual errors and mean CRLB loss for commonly used Zernike modes and optimized modes when imaging (i) point-like objects and (ii) extended volumetric objects. The error bars show the standard deviation of the randomly distributed initial aberration synthesized for the analyses and the residual aberration after correcting using methods of different biasing strategies. The insets are examples of aberration-free and noise-free images synthesized from the corresponding objects. The optimized biases used in (b) were derived in (a). Defocus has been omitted from the analysis for the volume object in (b)(ii).

## 3. Optimized bias strategy for fixed image exposure

In a practical imaging scenario, the AO process is likely to be limited by time, illumination power and total specimen exposure. A fair comparison between different biasing schemes must take into account these practical constraints. For this reason, we modelled and compared different sensorless AO correction processes in a 2-P microscope with a fixed total photon budget but a varying number of biases. The total photon budget was defined as a fixed illumination power over a fixed total exposure time during the correction process. For example, for a correction process that required ten measurements, each measurement had only 1/5 of the exposure time compared to another correction process requiring two measurements, if the total photon budget was kept the same. We selected three different levels of photon budget (high, medium and low) for analysis (for details, please see the supplementary information). A machine learning-based algorithm was used for the sensorless AO process (for details, please see the supplementary information). The initial aberration was generated consisting of random combinations of seven Zernike modes (astigmatism: $Z_5$, $Z_6$; coma: $Z_7$, $Z_8$; trefoil: $Z_9$, $Z_{10}$; spherical: $Z_{11}$) with the RMS value smaller than 3 rad.

We demonstrated how the number of input images $M$ can affect the machine learning-based sensorless AO accuracy when imaging at different photon budgets in Fig. 4. We chose three biasing strategies for comparison: 1) 2 biases, which has been shown in [31] to be the minimum required for MLAO operation; 2) $N+1$ biases, which is the minimum required to span a correction space of N coefficients [35]; 3) $2N$, which provides symmetrical biasing in each Zernike mode. These three models are logical subsets of all possible choices that are sufficient to show the trends behind different types of biasing schemes.

Like the results generated in section 2, Fisher information-based analysis was performed following the method section to derive for each case the optimized bias aberrations, which were shown as insets to the three subplots.

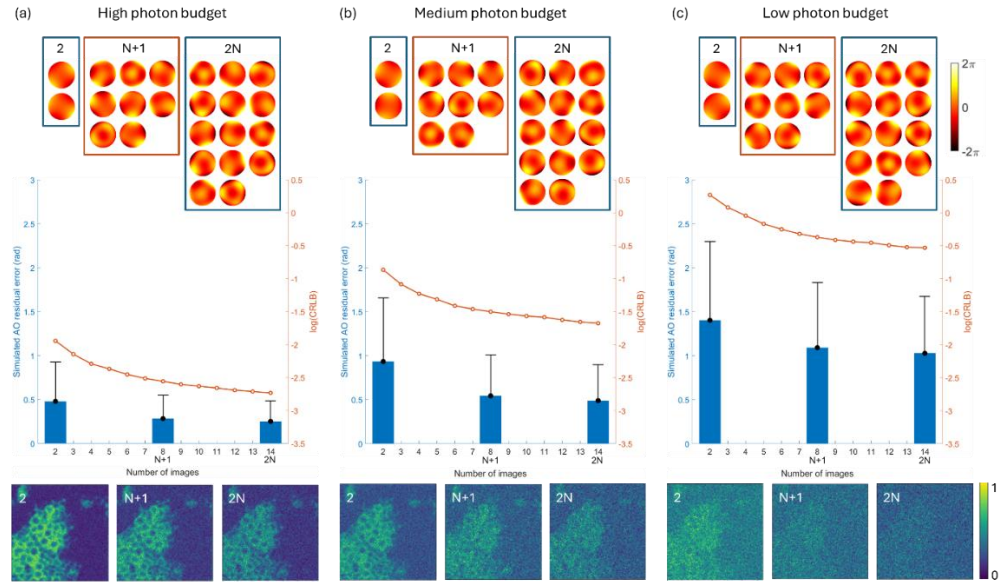

Fig. 4. (a-c) The mean CRLB loss on a logarithmic scale and simulated AO mean residual error with respect to the number of biases. Insets showed the optimal biases derived from Fisher information analysis. The images at the bottom are representative examples of typical input images to different AO algorithms when analyzed at different levels of photon budget. The images are displayed under the same color scale and the color-bar is of an arbitrary unit.

In all three imaging scenarios with different photon budgets, the sensorless AO algorithm with $2N$ biases always resulted in the best performance. Notably, the relative benefit in performance brought by a larger bias number became marginal as it grew.

The results have a clear heuristic interpretation. One way to understand the results is to imagine the algorithm with 2 biases as $N$ repeated measurements with the same 2 biases but $1/N$ of the photon budget for each measurement. The resultant algorithm is thus similar to the $2N$ algorithm, where each measurement has the same photon budget, but with sub-optimal biases (as $2N$ used the optimized biases derived from the Fisher information analysis). It was hence expected that under the same photon budget, the algorithm with 2 biases would have less accurate correction than the one with $2N$ biases and the performance was better as the number of biases increased.

An important observation is that CRLB correctly predicted the performance of different configurations of the sensorless AO algorithms. This indicates that the CRLB analysis could be used to guide the design of optimal bias aberrations for new sensorless AO schemes.

**Conclusion**

We used Fisher information analysis to optimize different sensorless AO methods for multiphoton microscopes in three practical applications. For the intensity-based $2N+1$ parabolic fitting algorithm, we optimized suitable bias amplitude for each bias mode which have close proximity to the minimum residual error. We derived the optimized bias modes when imaging different objects. For a point-like object, the resultant optimal modal combinations were close to a defocus shape and, for a volumetric object, were close to an astigmatism shape. This agreed with previously reported results [34]. We also discussed scenarios where different correction schemes were used while maintaining the same total photon budget. This permitted a fair practical comparison between having a longer single-image exposure time and fewer measurements, or to have a shorter exposure time and more measurements. Results suggested that more measurements provide better-conditioned phase information, although the additional benefit becomes more marginal when the number of measurements is increased.

We understand that Fisher information assumes no bias towards any provided information, which is not true in most practical sensorless AO method algorithms. For this reason, the information analysis should be used with care in conjunction with other prior knowledge about the imaging process. Discussions in previous work on MLAO [34] suggested that converged neural networks assign different weights to features of different spatial sizes, which, in return, reflects that different information may have different relevance to phase extraction. Furthermore, certain bias aberrations, such as defocus, may result a high Fisher information due to the axial shift of the focal plane and variations in the 3-D sample structures; however, such changes to the resultant images contain mainly information about the sample structure and do not necessarily lead to more effective estimation of other aberration modes. This is the reason why we excluded the defocus mode when analysing the imaging scenarios of the extended 3-D objects presented in Fig. 3b. Nevertheless, with careful design to the analysis, Fisher information remains as a complete measure and thus sets an upper bound for the obtainable information content.

The work demonstrated the great flexibility of the Fisher information analysis and provided a versatile framework for independently optimized, better informed, sensorless AO method design. In many practical situations, when direct comparisons between different algorithm settings are too time consuming and infeasible, Fisher information analysis provides a possible alternative offering a statistical and reliable overview of the method performance. In the future, variations to this analysis that compute more targeted aberration-related information can be implemented for more method-specific analyses. This work provides a strong information-guided approach that is complementary to the heuristic empirical experience-based method design.

**Funding**

This research was supported by UKRI/Wellcome Physics of Life grant EP/W024047/1, European Research Council Advanced Grant AdOMiS 695140, and Schmidt Sciences LLC.

**Acknowledgement**

The authors would like to thank Prof. Fang Huang's group from Purdue University for inspiring discussions about this work.

**Disclosures**

The authors declare no conflicts of interest.

**Data availability**

The datasets generated and/or analyzed during the current study are available from the corresponding author upon reasonable request.

**Supplementary information**

See supplementary information document for supporting content.

# Information-guided optimization of image-based sensorless adaptive optics methods: supplementary information

**BIWEI ZHANG, MARTIN J. BOOTH, AND QI HU**

*Department of Engineering Science, University of Oxford, Oxford OX1 3PJ, UK*

## 1. CRLB optimization

The optimized biases $a_k$, were initialized randomly where each followed a uniform distribution over the lower half range of the complete searching range. For example, if the total aberration for analysis had a root mean square (RMS) value not larger than 3 radians, and hence the highest sensible bias amplitude was also 3 radians, each $a_k$ were initialized randomly and uniformly over the range 0 to 1.5 radians.

The CRLB partial derivatives were computed numerically. The infinitesimal steps $\partial\theta$ were chosen to be 0.01 for the partial derivative computation.

The optimization was done in batches to avoid memory leakage. The batch size was chosen to be 16 when the number of input images is not larger than 8; otherwise, the batch size was chosen to be 8. There were 1,000 batches generated for training and 100 batches for validation.

The ADAM gradient descent algorithm was used for optimization. The parameters were updated after each batch. A variable learning rate was used. The initial learning rate was 0.05, and decreased by half every 200 epochs.

## 2. Zernike modes

Noll's index was used for Zernike modes throughout this paper [1]. A summary of appeared Zernike modes is provided in Table S1.

In Table S1, $Z_n^m(r,\theta)$ is defined in a unit circle with polar coordinates by

$$Z_n^m(r,\theta) = \begin{cases} m < 0: \sqrt{2} R_n^{-m}(r) \sin(-m\theta) \\ m = 0: R_n^0(r) \\ m > 0: \sqrt{2} R_n^m(r) \cos(m\theta) \end{cases},$$

$$R_n^m(r) = \sqrt{n+1} \sum_{s=0}^{(n-m)/2} \frac{(-1)^s (n-s)!}{s!\left(\frac{(n+m)}{2}-s\right)!\left(\frac{(n-m)}{2}-s\right)!} \tag{S1}$$

**Table S1. A summary of Zernike modes ($Z_k$) in Noll's index**

| k | n | m | $Z_n^m(r,\theta)$ | Name | Shape |
|---|---|---|---|---|---|
| 4 | 2 | 0 | $\sqrt{3}(2r^2 - 1)$ | Defocus | 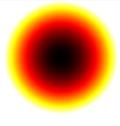 |
| 5 | 2 | -2 | $\sqrt{6}r^2 \sin(2\theta)$ | Oblique astigmatism | 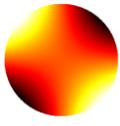 |
| 6 | 2 | 2 | $\sqrt{6}r^2 \cos(2\theta)$ | Vertical astigmatism | 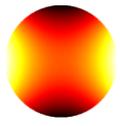 |
| 7 | 3 | -1 | $2\sqrt{2}(3r^3 - 2r) \sin(\theta)$ | Vertical coma | 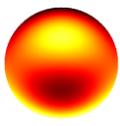 |
| 8 | 3 | 1 | $2\sqrt{2}(3r^3 - 2r) \cos(\theta)$ | Horizontal coma | 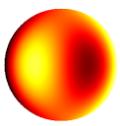 |
| 9 | 3 | -3 | $2\sqrt{2}r^3 \sin(3\theta)$ | Vertical trefoil | 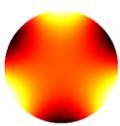 |
| 10 | 3 | 3 | $2\sqrt{2}r^3 \cos(3\theta)$ | Oblique trefoil | 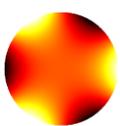 |
| 11 | 4 | 0 | $\sqrt{5}(6r^4 - 6r^2 + 1)$ | Primary spherical | 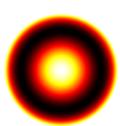 |

### 3. Photon budget

The photon budget was defined as the total signal photon number of the input image stack when aberration free (i.e. when Strehl ratio equals to 1). For the aberrated and biased images, the total signal photon number of the image was scaled by its Strehl ratio. The more measurements per input image stack, the smaller fraction of the photon budget was assigned to each input image of the stack. The size of the synthesized images was $128 \times 128$ pixels. For high/medium/low photon budget, the signal photon was chosen to be 1,000,000/320,000/100,000 respectively. In addition, a background offset was added to each image. A fixed photon number of 100 was added to each image stack (and hence a smaller fraction of the background photon number was added to the image stack with more measurements). The summed photon value of each pixel was the mean value following a Poisson distribution.

### 4. Machine learning-based algorithm

The machine learning based sensorless AO algorithm employed a ResNet-18 architecture [2]. The last layer of the NN was adjusted for different outputs. Thirty epochs were trained and the loss function was the root mean square error (RMSE) of the residual aberration. The batch size was 32. The learning rate was set as $10^{-3}$ and decayed to 1/3 of its previous value when plateaued (the decrease of the validation loss was less than $10^{-3}$) until the learning rate reached $10^{-5}$.

The NN was trained with 20,000 input-output dataset pairs where image stacks were the input and aberration coefficients were the output. The validation dataset size was 2,000. Both the training and validation dataset were synthesized as defined in Equation 7. The aberration was randomly generated from a n-sphere distribution [3] and its RMS value was between 0 and 3 rad. The synthesized image stack was then normalised such that the maximum pixel reading of the stack was 1 and the minimum pixel value was 0.

### 5. Modal ambiguity (relating to Results section 2 about coma and trefoil for point-like and volumetric objects)

In Results section 2, It was observed that Zernike modes with odd radial order ($n$) performed much worse as biases for point-like object comparing to volumetric objects. Table S2 below provides illustration of this phenomenon. The table shows sections through the PSF for different bias modes, in the presence of ±1 radian of an astigmatism input aberration, for both in-focus and out-of-focus planes.

Note that when an aberration consists of even radial order modes only (astigmatism $Z_5$ in this case), if the imaging plane is in-focus, the resultant PSFs of the same aberration but the opposite sign look identical upon an odd radial order bias (see column 2 and 3). For volumetric samples, due to the out-of-focus sample structures, the resultant PSFs can have distinct shapes to make the signs of the aberration distinguishable (see column 4 and 5).

**Table S2. A comparison of PSFs for different bias modes**

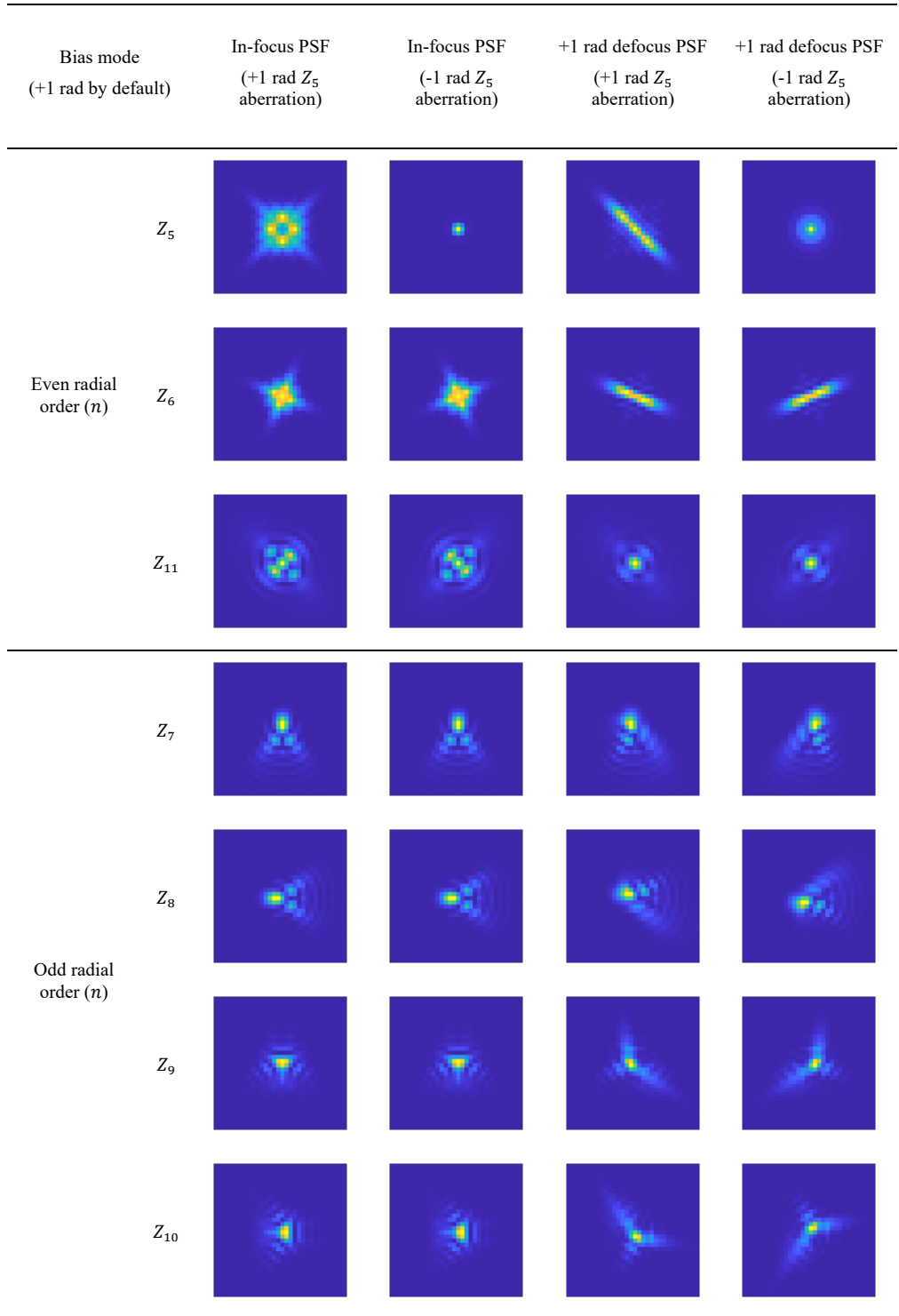

## 6. Optimized biases at different photon budgets

In Results section 3, we derived the optimized bias modes for different imaging scenarios under different photon budgets. In this section, we show in details of the optimized bias modes.

The RMS values of the derived biases were shown in Fig. S1 and were computed as

$$RMS = \sqrt{\sum_{l=5\ to\ 11} a_l^2}, \qquad (S2)$$

where $a_l$ were the Zernike modal coefficients for each bias. One interesting observation is that the RMS values of the biases generally increase when the number of biases for each correction cycle increases.

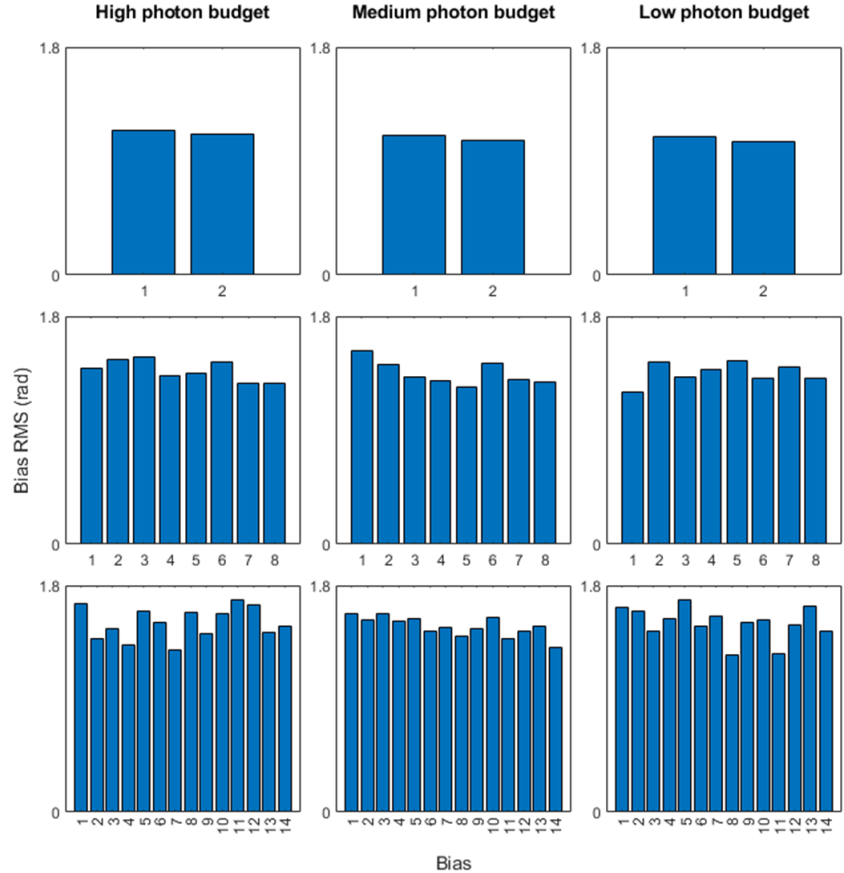

Fig. S1. RMS values of the derived optimized biases for each aberration correction scheme under different photon budget., for the 2 bias (top row), $N+1$ bias (middle row) and $2N$ bias (bottom row) configurations.

To show more clearly the modal composition of the biases, we grouped together modes of the same type. The reason for this was due to the arbitrary orientation of each mode group that could be generated by the optimization algorithm. The oblique and horizontal astigmatism modes 5 & 6 were grouped together; similarly grouped were the vertical and horizontal coma modes 7 & 8, as were the vertical and horizontal trefoil modes 9 & 10; the spherical mode 11 was considered on its own. The modal decompositions of each bias were shown in Fig. S2 and the percentage components were computed in terms of the mean square mode values as

$$\text{\% of the modal group} = \frac{\sum_{modes\ in\ group} a_l^2}{RMS^2}. \qquad (S3)$$

For the correction strategy with just two biases, astigmatism was the dominant component. When the number of biases increased, the component of trefoil increased. Coma modes remained small. For the correction strategy with $2N$ biases, there was no dominant modal shape and both coma and spherical featured significant contributions.

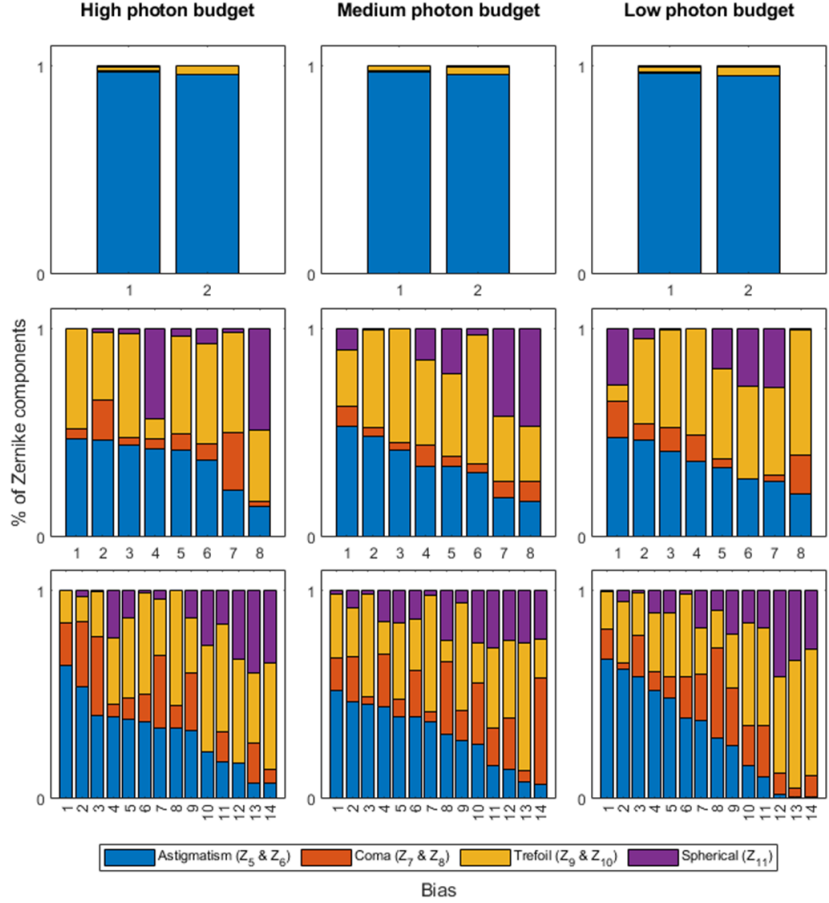

Fig. S2. Percentage of modal group components of the optimized biases for the 2 bias (top row), $N+1$ bias (middle row) and $2N$ bias (bottom row) configurations.

In addition, we computed the inner products between the bias modes to show their orthogonality. They were shown in Fig. S3 and were computed as

$$\text{inner product} = \sum_{l=5\ to\ 11} a'_l a''_l, \qquad (S4)$$

where $a'_l$ and $a''_l$ were the Zernike modal components for two biases. The bias modes generally were predominantly orthogonal as shown by the diagonal dominance of the matrices. This represents a close to even distribution of bias modes, providing good spanning of the aberration coefficient space.

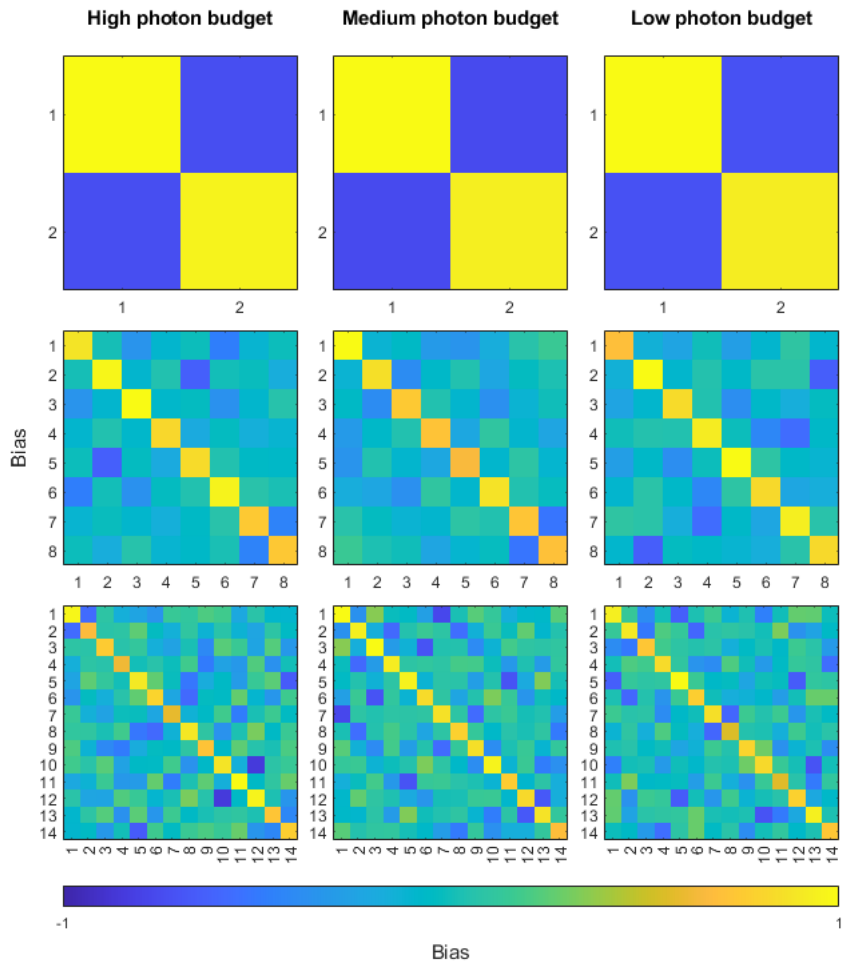

Fig. S3: Matrices of inner products computed among the optimized biases for the 2 bias (top row), $N+1$ bias (middle row) and $2N$ bias (bottom row) configurations.